\documentclass[smallcondensed]{svjour3}
\usepackage[italian, english]{babel}
\usepackage[latin1]{inputenc}
\usepackage[dvips]{graphicx}
\usepackage{color}

\journalname{Archive for History of Exact Sciences}

\begin{document}

\title{The quest for the size of the universe in early relativistic cosmology (1917-1930)}

\author{Giulio Peruzzi \and
        Matteo Realdi}

\institute{
Department of Physics, University of Padova\\
via Marzolo 8, 35121, Padova, Italy\\
\email{giulio.peruzzi@unipd.it; matteo.realdi@unipd.it}
}

\date{Received: date / Accepted: date}

\maketitle

\begin{abstract}

Before the discovery of the expanding universe, one of the challenges faced in early relativistic cosmology was the determination of the finite and constant curvature radius of space-time by using astronomical observations. Great interest in this specific question was shown by de Sitter, Silberstein, and Lundmark. Their ideas and methods for measuring the cosmic curvature radius, at that time interpreted as equivalent to the size of the universe, contributed to the development of the empirical approach to relativistic cosmology. Their works are a noteworthy example of the efforts made by modern cosmologists towards the understanding of the universe as a whole, its properties, and its content.

\keywords{
relativistic cosmology
\and curvature radius of space-time
\and expanding universe
\and Willem de Sitter
\and Ludwik Silberstein
\and Knut Lundmark
}

\end{abstract}

\section{Introduction}

A historical reconstruction of some attempts to determine the cosmic curvature radius by means of astronomical observations is presented in the following. The framework is that of the early phase of the science of the universe in the modern accepted meaning, from the beginning of relativistic cosmology in 1917 to the diffusion of theoretical models describing the expanding universe in 1930. Some leading scientists, who participated in the debate at that time, investigated the measurable properties of the universe, which was assumed to be static. In the light of the relativistic interpretation of the curvature of space-time, they speculated on the finite dimension of the universe, and, in this sense, attempted to extrapolate the size of the universe, i.e. the value of the cosmic curvature radius. After mentioning the very first suggestion in relativistic cosmology offered by Albert Einstein (1879-1955) on the value of the world radius, we focus on the analysis carried out by Willem de Sitter (1872-1934), who obtained a rough measure of the curvature radius of space-time. We then examine the different works in this field by Ludwik Silberstein (1872-1948) and Knut Lundmark (1889-1958). Finally, we give an overview of further investigations on the topic carried out before the discovery of the expanding universe. The variety of methods to obtain the curvature radius and corresponding estimates is summarized in table \ref{final}. The aim of this paper is to illustrate how the ideas, methods, and results proposed to measure the world radius can be read as noteworthy contributions to the first interplay between some of the speculative predictions of modern cosmology and the increasing amount of pertinent empirical evidence from the observable part of the universe.

The first two pioneering and rival cosmological solutions of relativistic field equations, formulated in 1917 by Einstein and by de Sitter, were the main focus of interest during the early phase of relativistic cosmology. These solutions represented by their intention a static universe with finite curvature radius. The Einstein universe was spherical and filled with matter, whereas the hyperboloidal (or equivalently hyperspherical) model of de Sitter was empty of matter and radiation. The possibility of extrapolating the radius of the universe (hereafter denoted by $R$) from observations was recognized as one of the interesting astronomical consequences of general relativity. However, such a quest represented an issue of secondary importance with respect to the investigation of the different properties of Einstein's and de Sitter's models in that period.

Actually, during the Twenties, the first link between theoretical cosmology and observational astronomy was mainly concerned with the interpretation of the displacement of wavelengths measured in spectra of stars, globular clusters, and especially spiral nebul{\ae}. The discovery of the expanding universe arose from the search for a suitable relation between observable quantities such as spectral shifts, distances, and apparent diameters. It was especially de Sitter's empty universe which attracted interest among scientists since it offered a (rather puzzling) interpretation of spectral displacements measured in stars and spiral nebul{\ae}. With regard to the content of the universe, Edwin Hubble (1889-1953) proved in 1925 the very existence of spiral nebul{\ae} as truly extragalactic stellar systems. The subsequent consensus about the status of spirals thus marked the change from the conception of a universe made of stars and nebul{\ae} to the picture of a universe filled with galaxies (nowadays, galaxy clusters and superclusters are regarded as the fundamental pieces which contribute to the matter content of the universe). The transition to the expanding universe took place when Hubble himself, in collaboration with Milton Humason (1891-1972), provided the empirical evidence that distant galaxies receded from each other: in 1929 Hubble confirmed that a linear relation, later known as the ``Hubble law'', existed between redshift and distance of extragalactic nebul{\ae}. From 1930 on, such a discovery subsequently allowed the acceptance and the diffusion of the relativistic non-static and non-empty models of the universe, which had been formulated already in 1922 by Aleksandr Friedmann (1888-1925), and independently in 1927 by Georges Lema\^{i}tre (1894-1966). Eventually, it was in 1930 that the cosmological interpretation of redshift in spirals as due to the expansion of the universe, first proposed by Lema\^{i}tre in 1927, officially entered modern cosmology.

After the discovery of the expanding universe, the concept of a finite and constant world radius was superseded by the notion of the curvature radius depending on time $R(t)$, which later evolved to the present notion of the time-dependent expansion parameter $a(t)$, also called cosmic scale factor\footnote{In modern cosmology the expansion parameter $a(t)$ is related to the Gaussian curvature $C_{G}=\frac{k}{a^{2}}$. The parameter $k$ determines the constant curvature of spatial sections. It can be negative ($k=-1$), null ($k=0$), or positive ($k=+1$), yielding respectively an open universe (3-dimensional hyperbolical space), a flat universe (Euclidean space) or a closed universe (3-dimensional spherical space). In fact, the curvature parameter can be scaled in such a way to assume only the values $k$ = (1, 0, -1).  The parameter $a(t)$ thus represents the radius of spatial curvature, which in cosmology describes the modulus of Gaussian curvature radius $R_{G}=C_{G}^{-1/2}=\frac{a}{\sqrt{|k|}}$ \cite[pp. 9-13]{Coles-Lucchin 2002}.}.

In this picture, the efforts made in the period 1917-1930 to specify the spatial extent of the universe represent a short but significant chapter in the history of the early development of relativistic cosmology. The present paper intends to show the intrinsic interest and the reactions that the quest for the size of the universe stimulated in those years\footnote{The present work is mainly based on chapters 5 and 6 of: Realdi, Matteo. 2009. \textit{Cosmology at the turning point of relativity revolution. The debates during the 1920s on the `de Sitter effect'} (PhD Thesis, University of Padova), from which some parts have been taken, and here adapted.}. De Sitter, Silberstein, and Lundmark ventured the path, now in the new framework of relativistic cosmology, of directly estimating $R$ by using several astronomical objects, which, by their intention, played the role of distance indicators. Going beyond Einstein's first attempt, these authors used either observations of stars, or assumptions on the mean density of matter, or velocities of globular clusters and spiral nebul{\ae} in order to determine the value of $R$. Indeed the study of distant galaxies during the Twenties marked a turning point in the empirical approach to cosmology. In this context, the use of distance indicators such as stars and globular clusters was soon after recognized as not relevant to search for the curvature of space and to distinguish between cosmological models. Nevertheless, the attempts made by de Sitter, Silberstein, and Lundmark influenced the early debate about the observational tests of the first relativistic world models. Moreover, such attempts can be considered as valuable aspects of the broader scientific aim to interpret empirical evidence on astronomical scales through the laws of physics.

The analysis given in the following pages will highlight the different approaches and the significance such authors attached to the science of the universe. On the one hand, we shall see, de Sitter inaugurated in his works the \textit{systematic} attempt to relate astronomical observations from available data to the geometry of the entire universe in the framework of general relativity. He studied different methods to determine the world radius of both Einstein's universe and of his empty model. Nonetheless, de Sitter clearly argued for the very speculative meaning of investigating the universe as a whole, claiming that all conclusions drawn beyond observations had to be considered pure extrapolations. On the other hand, Silberstein addressed in 1924 the question of the determination of the curvature radius of de Sitter's space-time by means of globular clusters. He firmly denied the general cosmic recession of test particles in de Sitter's universe, which had been predicted in 1923 both by Arthur Eddington (1882-1944) and independently by Hermann Weyl (1885-1955). Silberstein formulated a theoretical linear relation between shift and distance, which he applied to the observed receding and approaching motions of globulars. Lundmark showed that the result obtained by Silberstein was not correct. In his detailed empirical analysis on some classes of stars and spiral nebul{\ae}, Lundmark proposed several values of the curvature radius of de Sitter's universe, and, in agreement with the cosmology of Carl Charlier (1862-1934), suggested the picture of a hierarchical distribution of stars and nebul{\ae}.

With its focus on the specific question of early attempts to estimate the value of the cosmic curvature radius before the discovery of the expanding universe, the present analysis offers an additional point of view supplementing those given in the vast scientific literature, which already exists on the history of the early developments of modern cosmology (see for instance \cite{Ellis 1989,Ellis 1990,Hetherington 1996,Kerszberg 1989,Kragh-Smith 2003,North 1965,Nussbaumer-Bieri 2009,Osterbrock 1990,Smith 1982,Smith 2009}).

\section{Einstein, the cosmological constant, and the curvature radius}

In 1917, as mentioned above, Einstein and de Sitter proposed two different cosmological solutions of relativistic field equations. Incidentally, it was in the same year that the Hooker 100-inch (2.5 m) telescope, the instrument which was used by Hubble during the following years for his capital contributions to observational cosmology, saw ``first light'' on Mount Wilson, California\footnote{For a historical account of the Mount Wilson Observatory, see \cite{Sandage 2004}.}.

The origin of inertia can be viewed as the main question, which actually led Einstein and de Sitter to the formulation of their respective models of the universe\footnote{The debate between Einstein and de Sitter, which marked the beginning of relativistic cosmology, is analyzed in: \textit{The Einstein - de Sitter - Weyl - Klein debate}, in \cite[pp. 351-357]{CPAE 1998}. The authors of the present paper reconstructed part of the Einstein - de Sitter correspondence in \cite{Realdi-Peruzzi 2009}.}. In his famous paper \textit{Kosmologische Betrachtungen zur allgemeinen Relativit\"{a}tstheorie} (\textit{Cosmological considerations in the general theory of relativity}), which appeared in February 1917, Einstein proposed a finite and unbounded universe, where the spatial sections at constant time were spherical, with constant positive curvature radius $R$. Accounting for the complete material origin of inertia, as inspired by some ideas of Ernst Mach (1838-1916), such a closed universe involved that inertia was uniquely determined by the interaction between masses. In this way, Einstein achieved what he called the relativity of inertia, i.e. he avoided the necessity of assuming that any independent property of space could be claimed to be at the origin of inertia. Furthermore, by means of the spatial closedness he overcame the difficulty of obtaining values of the gravitational potentials (identified by the symbols $g_{\mu\nu}$) which at infinity were invariant for all transformations.

Einstein disregarded local non-homogeneous distributions of matter (like stars and planets), and introduced in his model an extremely small density of matter which hypothetically was uniformly and homogeneously distributed through space. It is worth noting that the assumption of global average properties of matter, when considering cosmological scales, turned out to be one of the typical features in the modern approach to cosmology. The condition of the homogeneity and isotropy is now referred in the literature as the ``cosmological principle'': matter and radiation are assumed to be uniformly distributed through space on very large scales, with neither privileged directions, nor privileged positions. In the present picture of the expanding universe, the cosmological principle asserts that the universe exhibits the same properties at any given cosmic time, i.e. that the 3-space surfaces of constant cosmic time are homogeneous and isotropic.

Einstein introduced in his field equations the so-called cosmological term, $\lambda g_{\mu\nu}$, which, by his intention, acted like an anti-gravity term and accounted for the supposed static equilibrium of the universe. At the beginning of the last century, in fact, astronomical observations did not (yet) reveal any large-scale systematic velocity fields. With the cosmological constant, the relativistic field equations, which relate the space-time geometry (on the left-hand side) to the energy-matter content of space-time (on the right-hand side), took the form:
\begin{equation}
G_{\mu\nu}-\frac{1}{2}g_{\mu\nu}\mathcal{G}-\lambda g_{\mu\nu}=-\kappa
T_{\mu\nu}.
\end{equation}
Here $G_{\mu\nu}$ is the Ricci tensor, i.e. the contracted Riemann
curvature tensor, $\mathcal{G}$ is the Ricci scalar, i.e. the scalar curvature obtained
from $\sum\,g^{\mu\nu}G_{\mu\nu}$, and $T_{\mu\nu}$ is the energy-momentum
tensor; $\kappa$ is a constant equal to $\frac{8\pi\\G}{c^{4}}$, where $c$ is the speed of light and $G$ is the gravitational constant. In this way, the metric of a static and closed universe now resulted as a coherent solution of modified field equations. This was the ``rough and winding road'' \cite[p. 423]{Einstein 1917}, which Einstein acknowledged he followed in order to show that the theory of general relativity led to a system free of contradictions.

The value of the cosmological constant $\lambda$ was related to $R$, and to the mean density
of world matter $\rho$ \cite[p. 431]{Einstein 1917}:
\begin{equation}
\lambda=\frac{1}{R^{2}}=\frac{\kappa\rho c^{2}}{2}\,.
\end{equation}
By means of such a relation, Einstein derived the first estimate of $R$ in the framework of relativistic cosmology. This estimate was not reported in his 1917 cosmological paper, but can be found in some correspondence \cite[docs. 298, 300, 306, 308, 311]{CPAE 1998}. In these letters, Einstein reported that, from star counts, the spatial density of matter was of the order of $\rho\simeq 10^{-22}$ g/cm$^{3}$. Therefore, from the previous relation the world radius of his model was $R=10^{7}$ light-years ($6\cdot10^{11}$ AU)\footnote{The astronomical unit (AU) and the light-year are units of distance in astronomy and cosmology. The astronomical unit corresponds to nearly $1.49\cdot10^{11}$ m, whereas the light-year is approximately $9.42\cdot10^{15}$ m, or roughly 63,272 AU. Another unit of distance is the parsec (pc), equal to $3.09\cdot10^{16}$ m (1 pc = 206,264.8 AU $\simeq$ 3.26 light-years).}. The farthest visible stars, Einstein remarked, were estimated at $10^{4}$ light-years ($6\cdot10^{8}$ AU).

Later on, Einstein showed interest in the possibility of determining the value of the cosmological constant. Although such an estimate would directly reveal the value of the curvature radius, Einstein was mainly interested in the confirmation of the \textit{existence} in nature of the cosmological constant, which, by quoting Einstein himself, ``is not justified by our actual knowledge of gravitation'' \cite[p. 432]{Einstein 1917}, and which he had only introduced in 1917 to account for a static universe. A hint about this question can be found in some considerations which Einstein proposed in 1921 on the application of the Newtonian law of gravitation to globular clusters, and on the stationary equilibrium of such stellar systems \cite[pp. 394-395]{Einstein 1921a}, \cite{Einstein 1921b}. In this framework, Einstein referred to some observations made by Erwin Freundlich\footnote{We refer to \cite{Crelinsten 2006} for a historical reconstruction of the attempts made by astronomers to test Einstein's theory of relativity; see \cite{Hentschel 1994} for the role played by Freundlich in testing relativity theory.} (1885-1964). In fact, the comparison between the observed stellar velocities and the average theoretical velocity of stars (the latter obtained through the virial theorem in the case of Newtonian forces) would have possibly revealed the presence of a non-zero cosmological constant, which acted like an anti-gravity, and was thus able to keep the equilibrium of globular clusters. However, such an attempt, which is thoroughly analyzed in the related notes of \cite{Einstein 1921b}, did not give reliable results\footnote{In 1931, confronted with the empirical evidence that galaxies were receding from each another, Einstein abandoned the cosmological constant which he had introduced in 1917 in order to express in general relativity the static nature of the universe. In fact, this hypothesis was contradicted by the observed recession of spiral nebul{\ae} which now supported the interpretation of the expanding universe \cite{Einstein 1931}. For a historical account of the cosmological constant, see \cite{Earman 2001}.}.

\section{The size of the universe according to de Sitter}

The second relativistic model of the universe was proposed by de Sitter right after Einstein's 1917 cosmological paper was published. As discussed above, in Einstein's universe the world matter, uniformly distributed through space, was the only thing responsible for the origin of inertia, and the cosmological constant was necessary for a static world model. On the contrary, in his own model, de Sitter assumed that the contribution of the density of matter could be disregarded on the largest scales. The Dutch astronomer retained in his solution the $\lambda$-term, which was now responsible for the curvature of space-time and for the inertia of a hypothetical test particle inserted in such a world free of matter.

In fact, de Sitter developed a suggestion by Paul Ehrenfest (1880-1933), and extended Einstein's hypothesis of a finite 3-dimensional space to a finite 4-dimensional space-time of positive constant curvature. This model corresponded to a hypersphere embedded in Euclidean space, or, equivalently, to a hyperboloid embedded in Minkowski space-time. In such a model, the curvature radius was related to the cosmological constant through the relation \cite[doc. 313]{CPAE 1998}:
\begin{equation}
\lambda=\frac{3}{R^{2}}\,.
\end{equation}
De Sitter took into account the cosmological term in order to satisfy what he called the \textit{mathematical} postulate that at infinity the potentials were invariant under all transformations, a postulate which, according to de Sitter himself, did not have any real \textit{physical} meaning. It is for this purpose that he proposed a cosmological solution where all $g_{\mu\nu}$ were zero, i.e. an empty model of the universe. As de Sitter wrote in 1932, the cosmological constant was ``a name without any meaning, which (...) appeared to have something to do with the constitution of the universe; but it must not be inferred that, since we have given it a name, we know what it means. (...) It is put in the equations in order to give them the greatest possible degree of mathematical generality'' \cite[p. 121]{de Sitter 1932b}. On the one hand, as a matter of fact, de Sitter clearly acknowledged the importance of the \textit{mathematical} solutions that general relativity now offered for studying the universe as a whole. On the other hand, however, he pointed out that such investigations were built upon pure \textit{hypothesis} which could never be proven by empirical evidence since they referred to unobservable parts of the universe, and thus corresponded to \textit{extrapolations} beyond our neighborhood which ``can not be decided by physical arguments, but must depend on metaphysical or philosophical considerations'' \cite[p. 1222]{de Sitter 1917a}.

An interesting hint of this feature of de Sitter's approach to the cosmological question can be found in the original manuscripts written by the Dutch astronomer at the very early stages of relativistic cosmology. We mention for instance some notes on general relativity reported by de Sitter after conversations in Leiden between Einstein, de Sitter himself, Ehrenfest, and Gunnar Nordstr\"{o}m (1881-1923) on September 28-29, 1916. Actually, this first exchange on the problem of boundary conditions (which some months later Einstein solved by introducing a finite universe, whereas de Sitter solved it by considering that all potentials at infinity should be zero) marked the beginning of the debate between Einstein and de Sitter on the relativistic description of the universe as a whole. De Sitter wrote in late 1916 that:
\begin{quote}
Einstein wants the \underline{hypothesis of the closedness} of the world. He means by that that he makes the \underline{hypothesis} (conscious that it is a hypothesis which cannot be proven) that at infinity (that is at very large, \underline{mathematically} finite, distance, but further than any observable material object (...)) there are such masses (...) that the $g_{\mu\nu}$ at infinity assume \underline{certain degenerate values} (these have not to be 0, that is a priori not to be said), \underline{the same} in \underline{all} systems. (...) He is even prepared to give up the complete freedom of transformation (...). If it is possible to find a set of degenerate values of the $g_{\mu\nu}$, that is invariant for a not too restricted group of transformations, is a question that can be solved mathematically. Is the answer \underline{no} (what Ehrenfest and I expect), then Einstein's hypothesis of the closedness is untrue. Is the answer yes, then the hypothesis is not incompatible with the relativity theory. However, I \underline{even then} maintain my opinion that it is incompatible with the \underline{spirit} of the principle of relativity. And Einstein admits that I have the right to do so. Also the rejection of the hypothesis is completely admissible in the relativity theory\footnote{``Einstein wil de \underline{hypothese van de afgeslotenheid} der wereld. Hij verstaat daaronder dat hij de \underline{hypothese} maakt (bewust dat het een onbewijsbare hypothese is) dat er in het oneindige (d.i. op zeer groote, \underline{mathematisch} eindige, afstand, maar verder dan eenig waarneembaar materieel object, (...)) zoodanige massa's zijn (...) dat de $g_{ij}$ in het oneindige \underline{bepaalde ontaarde waarden} aannemen (deze hoeven niet 0 te zijn, dat is a priori niet te zeggen), \underline{dezelfde} in \underline{alle} co\"{o}rdinatensystemen. (...) Hij is dan ook bereid de volkomen vrijheid van transformaties op te geven (...). Of het mogelijk is een stel ontaarde waarden der $g_{ij}$ te vinden die invariant is voor een niet te erg beperkte groep van transformaties, is een vraag die mathematisch uit te maken is. Luidt het antwoord \underline{neen} (wat Ehrenfest en ik verwachten) dan is Einstein's hypothese der afgeslotenheid onwaar. Luidt het antwoord ja, dan is de hypothese niet in strijd met de relativiteitstheorie. Evenwel houd ik \underline{ook dan} vol dat zij wel in strijd is met den \underline{geest} van het relativiteitsprincipe. En Einstein geeft toe dat ik daartoe het recht heb. Ook verwerping der hypothese is volkomen geoorloofd in de relativiteitstheorie''. English translation by Jan Guichelaar.} \cite[AFA-FC-WdS-132]{de Sitter Archive}.
\end{quote}
In addition to these considerations, some 1917 correspondence with Jacobus Kapteyn (1851-1922), whom de Sitter asked for advice right during his debate with Einstein, reveals the great importance de Sitter attached to the actual picture drawn by the astronomical investigation of the observable part of the universe, which, as we shall see later, was helpful when comparing the different astronomical consequences between his own model and Einstein's. De Sitter was interested in the most recent advances on parallax measures and estimates of the total mass of the Milky Way. Furthermore, he paid special attention to whether observations of stars and nebul{\ae} revealed a nearly \textit{systematic} redshift. ``These are hard nuts - Kapteyn replied to de Sitter in June 1917 - you are giving me to crack'' \cite[p. 96]{van der Kruit-van Berkel 2000}.

In his papers de Sitter studied the properties of Einstein's universe, which he labeled as ``system A'', and those of his own empty model, labeled as ``system B''. He took into account for the sake of comparison also the properties of the solution of field equations without $\lambda$, i.e. of the line element of the special theory of relativity (``system C''). De Sitter proposed several forms of the line element of his cosmological solution, which depended on the choice of coordinates. In one of these, which became known as the ``static form'' of de Sitter's universe, the metric was the same as in Einstein's spherical model, except for the potential term which referred to the time coordinate $t$. In Einstein's world such a potential was $g_{tt}=1$, whereas in the static form of de Sitter's universe such a potential term depended on the variable $r$ (interpreted as a radial coordinate) and on the radius $R$ \cite[p. 230]{de Sitter 1917b}:
\begin{equation}
g_{tt}=\cos^{2}\frac{r}{R}.
\end{equation}
Rather than a spherical space, de Sitter assumed for his universe that the physical (closed) space was elliptical. Such a choice, in fact, avoided the existence of the antipodal point which one observes in the spherical space, where any two geodesics intersect each other in two points. To imagine the difference between these two spaces, one can consider on the one hand the analogy between spherical space and the surface of a sphere, on the other hand the analogy between elliptical space and the surface of a hemisphere \cite[p. 202]{Harrison 2000}. The volume of the elliptical space ($\pi\,R^{3}$) is half of the volume of the spherical one, and both of these spaces can be approximated to the Euclidean space for small values of $r$ compared to $R$, the curvature radius\footnote{By projecting the elliptical space on the Euclidean one through the transformation of coordinates $\mathbf{r}=R\,\tan\frac{r}{R}$, the potential term of the time coordinate became in this case:
$g_{tt}=\left(1+\frac{\mathbf{r}^{2}}{R^{2}}\right)^{-1}$ \cite[p. 232]{de Sitter 1917b}. Here the symbol $\mathbf{r}$, used by de Sitter in his papers, represents a spatial coordinate, not a vector. }.

It is clear that in the framework related to the investigation of the astronomical consequences of general relativity, de Sitter showed great interest in the determination of the curvature radius of space-time by means of observational evidence.

The first difference between the two relativistic cosmological models A and B emerged by considering trigonometric parallax\footnote{In system B the trigonometric parallax $p$ of a star at the distance $r$ from the Sun was:
$p\simeq\frac{a}{R\,\sin\frac{r}{R}}=\frac{a}{\mathbf{r}}\sqrt{1+\frac{\mathbf{r}^{2}}{R^{2}}}$,
$a$ being the average distance between the Sun and the Earth. Thus here the parallax $p$ was never zero, and reached
its minimum value at the distance from the Sun $r=\frac{\pi}{2}\,R$ \cite[p. 13]{de Sitter
1917c}. In system A such a parallax was $p\simeq\frac{a}{R}\cot\frac{r}{R}=\frac{a}{\mathbf{r}}$,
so that $p$ diminished to zero as the distance increased up to $r=\frac{\pi}{2}\,R$, and for larger distances it became negative \cite[p. 233]{de Sitter 1917b}.}. In the framework of the use of parallax to measure the curvature of space, a lower limit of the curvature radius, $R>4\cdot10^{6}$ AU, had been proposed already in 1900 by Karl Schwarzschild (1873-1916). In fact, in his pre-relativistic analysis on curved spaces\footnote{See \cite{Schemmel 2005} for an analysis of the work of Schwarzschild on the measure of the curvature of space by means of parallax.}, Schwarzschild had assumed that some stars had trigonometric annual parallax of the order of $p=0''.05$ \cite[p. 2541]{Schwarzschild 1900}. As confirmed by Kapteyn in a letter to de Sitter, such a limit for annual parallax was still in 1917 the most suitable result obtained by direct observations \cite[p. 96]{van der Kruit-van Berkel 2000}. As remarked by de Sitter:
\begin{quote}
The meaning [of observed parallaxes] is of course actually measured parallaxes, not parallaxes derived by the formula $p=a/r$ from a distance which is determined from other sources (comparison of radial and transversal velocity, absolute magnitude, etc.). Schwarzschild assumes that there are certainly stars having a parallax of $0''.05$. All parallaxes measured since then [1900] are \textit{relative} parallaxes, and consequently we must at the present time still use the same limit \cite[p. 234]{de Sitter 1917b}.
\end{quote}

De Sitter proposed alternative ways to determine the cosmic curvature radius.
With regard to Einstein's spherical universe, an estimate of $R$ was obtained by means of the relation between the angular (apparent) diameter $\delta$ of
an object of linear (absolute) diameter $d$ at the distance $r$ from the Sun.
``It is very probable - de Sitter wrote in 1917 -  that at least
some of the spiral nebul{\ae} or globular clusters are galactic
systems comparable with our own in size'' \cite[p. 24]{de Sitter 1917c}. By taking, for example, for these systems $d=10^{9}$ AU and
$\delta=5'$, de Sitter asserted that the radius of system A was $R\geq10^{12}$ AU.

Furthermore, the size of Einstein's universe was obtained by some assumptions on the density of world matter $\rho$, i.e. by applying the same method first suggested by Einstein \cite[p. 24]{de Sitter 1917c}. If such a
density was assumed to be the same as the star density at the center
of the galactic system (80 stars/1000 pc$^{3}$, or $\rho=10^{-17}$ g/cm$^{3}$), then
the curvature radius of system A resulted as $R=9\cdot10^{11}$
AU, not so different from the estimate suggested some months before by Einstein. From assumptions on
the number of galactic systems ($7\cdot10^{6}$), on their average shortest distance ($10^{10}$ AU), and on the total mass of the world matter sufficient to fill the whole universe with galaxies, the density resulted to be $\rho=\frac{1}{3}\cdot10^{-20}$ g/cm$^{3}$, so that the radius resulted: $R\leq5\cdot10^{13}$ AU.

In addition, as mentioned above, one
should expect in system A to see the antipodal image of the Sun.
Since this was not observed, light should have been
absorbed in its travel around the world. According to de Sitter, an absorption of 40
magnitudes, which had been proposed already in 1900 by
Schwarzschild, was sufficient to produce such an effect along the distance $\pi\,R$ of the spherical space.
Therefore, by assuming the result
by Harlow Shapley (1885-1972) of an absorption of 0.0001 mag/10 pc, de Sitter obtained the value
$R>\frac{1}{4}\cdot10^{12}$ AU for the curvature radius of
Einstein's universe \cite[p. 25]{de Sitter 1917c}.

With regard to the empty universe, estimates of $R$ could not be obtained by using the fact
that the ``back of the Sun'' was not observed. In fact, at the distance $r=\frac{\pi}{2}\,R$, i.e. at the largest possible distance in elliptical space, it was $g_{tt}=0$. Therefore, as de Sitter pointed out,
in such a system ``light requires an infinite time for the voyage round the
world'' \cite[p. 26]{de Sitter 1917c}. Moreover, the relation between
apparent and linear diameter of spirals could not be applied in system B, since at this distance the apparent angular diameter $\delta$ was also zero.

However, the empty universe showed an interesting feature, allowing de Sitter to propose a value of the curvature radius of system B by means of the displacement of wavelengths measured in spectra of some stars and spiral nebul{\ae}.
Since in the static form of system B the $g_{tt}$ potential term diminished with increasing $r$,
the frequency of light diminished
with increasing distances from the observer at rest at the origin of
coordinates: ``the lines in the spectra of
very distant objects - de Sitter wrote - must appear displaced towards the red''
\cite[p. 235]{de Sitter 1917b}.

As usual in those years, the spectral shift $z$ was mainly interpreted as a Doppler motion, i.e. as originated by relative motions through space between the observer and the observed object. The classic Doppler formula, used when the relative velocity $v$ is small when compared to speed of light $c$, is:
\begin{equation}
z\equiv\frac{\lambda_{o}-\lambda_{e}}{\lambda_{e}}=\frac{v}{c},
\end{equation}
where $\lambda_{o}$ and $\lambda_{e}$ denote, respectively, the wavelength measured
by the observer and the one emitted from the source. The redshift (or, respectively, blueshift) is due to an
increasing (decreasing) of wavelength, and is interpreted as a receding (approaching) motion of the source, with velocity $v$.

One of the relevant phenomena described in
Einstein's new theory of gravitation was the gravitational shift, i.e. the
contribution to the shift of spectral lines produced by stars themselves. In fact,
spectral lines originating in a strong
gravitational field, for example on the star's surface, would be
displaced towards the red to an observer in a weaker gravitational
field\footnote{For a star of mass $M_{\star}$ and density $\rho_{\star}$
(with solar mass and solar density $M_{\odot}=\rho_{\odot}=1$), such a
gravitational contribution to the measured spectral shift was equal to $0.634\,M_{\star}^{2/3}\,\rho_{\star}^{1/3}$ \cite[p. 719]{de Sitter 1916}.}.
In his analysis of the gravitational shift de Sitter referred to some observations by William Campbell (1862-1938). Already in 1911 Campbell found that a constant $K$ term, initially interpreted as a systematic error, had to be taken into account in order to determine the solar motion from stellar velocities of 35 groups of $B$ stars. ``An error of obscure source - Campbell wrote in 1911 - causes the radial velocities of Class $B$ stars to be observed too great by a quantity, $K$, amounting to several kilometers'' \cite[p. 105]{Campbell 1911}. As suggested by Kapteyn to de Sitter, such an average systematic shift to the red for $B$ stars corresponded to a receding velocity of the order of $v=+ 4.3$ km/sec \cite[p. 97]{van der Kruit-van Berkel 2000}, up to nearly $v=+ 4.5$ km/sec \cite[p. 719]{de Sitter 1916}.  De Sitter assumed that almost one third of this shift could be interpreted as the proper gravitational shift due to the source, i.e. at the star's surface. The remaining $v=+ 3$ km/sec corresponded to a ``spurious positive radial velocity'', in the sense that such a displacement of spectral lines was produced by the inertial field in system B, and in particular by the diminution of $g_{tt}$ in de Sitter's universe \cite[p. 26]{de Sitter 1917c}. Then $R$ could be obtained by considering the field produced by a fixed star\footnote{The relation between $R$, the star's distance $r$, and the star's spurious velocity $v$ (obtained by means of the Doppler formula) was in this case: $g_{tt}=\cos^{2}\frac{r}{R}\simeq\,1-2\frac{v}{c}=1-2\cdot10^{-5}$ \cite[p. 235]{de Sitter 1917b}.}. Since the average distance of $B$ stars was $r=3\cdot10^{7}$ AU, the curvature radius of the empty model resulted to be $R=\frac{2}{3}\cdot10^{10}$ AU \cite[p. 27]{de Sitter 1917c}.

De Sitter further developed such an analysis in relation to some nebul{\ae}.
Referring to data from the 1917 Council of the Royal Astronomical
Society \cite{Eddington 1917}, de Sitter
pointed out that the Small Magellanic Cloud was
estimated to be at $r>6\cdot10^{9}$ AU, with a radial
velocity $v\simeq+150$ km/sec. Consequently the
curvature radius of system B was $R>2\cdot10^{11}$ AU \cite[p. 27]{de
Sitter 1917c}. By some independent observations, three
nebul{\ae} (NGC 4594, NGC 1068 and the Andromeda nebula) showed
very large radial velocities compared with usual velocities of stars
in solar neighborhood (see table \ref{desittervel}). Taking $v=+600$ km/sec as the mean of their radial velocities, and the
curvature radius as $R=\frac{2}{3}\cdot10^{10}$ AU, the minimum distance of these
nebul{\ae} was $r=4\cdot10^{8}$ AU \cite[p. 236]{de
Sitter 1917b}. Alternatively, by assuming for these objects
a mean distance of about $r=2\cdot10^{10}$ AU, the
curvature radius of system B was
$R=3\cdot10^{11}$ AU \cite[p. 28]{de Sitter 1917c}.

\begin{table}

\caption{Radial velocities of the nebul{\ae} studied by de Sitter in 1917 \cite[p. 236]{de Sitter 1917b}. The average values which de Sitter used for his calculation are listed in the last column \cite[p. 27]{de Sitter 1917c}.}\label{desittervel}

\begin{center}
\begin{tabular}{llll}
\hline
 nebula            & observer  & velocity & average\\
                   &            & (km/sec)  & (km/sec) \\
\hline
 NGC 4594          & Pease     & +1180  &  +1185\\
                   & Slipher   & +1190          \\

 NGC 1068          & Pease     &  +765  &   +925\\
                   & Slipher   & +1100          \\
                   & Moore     &  +910          \\

 Andromeda         & Wright    & --304  &  --311\\
                   & Pease     & --329          \\
                   & Slipher   & --300          \\
\hline
\end{tabular}
\end{center}
\end{table}

De Sitter remarked that the estimates of the curvature radius of system A were very uncertain. However, he also acknowledged the scarcer value of the estimates of the size of his own universe obtained through the radial velocities of nebul{\ae}. Nevertheless, in face of the different physical consequences of these world models, he suggested that the discovery in the future of a systematic receding radial motion of spirals would allow a discrimination between system A and system B. His model, as a matter of fact, had the great advantage that it required an apparent positive radial velocity for distant objects, and the empirical confirmation of such a recession would suggest adopting model B rather than A. This was the concluding remark which de Sitter wrote in 1920, when he noted that now, thanks to the work of Vesto Slipher (1875-1969), the radial velocities of 25 spiral nebul{\ae} were known, showing a mean receding motion of $v=+ 560$ km/sec. However, de Sitter added that ``the decision between these two systems must, I fear, for a long time be left to personal predilection'' \cite[p. 868]{de Sitter 1920}.

The Dutch astronomer, who was the director of the Leiden Observatory from 1919 to 1934, did not deal with the cosmological question in other published papers up to 1930, and did not participate in the cosmological debates which took place during the Twenties. By quoting Eddington, de Sitter was ``the man who discovered a universe and forgot about it'' \cite[p. 925]{Eddington 1934}. Nevertheless, further considerations by the Dutch astronomer on the determination of the size of the universe can be found in some correspondence of the late twenties between him and Frank Schlesinger (1871-1943). In December 1929, now in view of Hubble's confirmation of the existence of a linear redshift-distance relation among galaxies, de Sitter wrote to Schlesinger that:
\begin{quote}
Now there comes a most unexpected and curious coincidence. If we suppose the whole world to be filled with spiral nebul{\ae} (...), the density becomes $2\cdot10^{-28}$ in c.g.s. units, which on first sight be thought not to differ much from emptiness. But if we now take Einstein's own solution A of his field equations, i.e. the solution for a \underline{full} world, then there is a relation between the radius of the universe and the density, and taking for the density $2\cdot10^{-28}$, the radius is found to be $2\cdot\frac{1}{4}\cdot10^{9}$ light-years ($3\cdot10^{13}$ AU), i.e. practically the same as for the empty world\footnote{It is interesting to note the remarkable difference between the estimate of the density of world matter suggested in 1917 by Einstein, which was $\rho\simeq 10^{-22}$ g/cm$^{3}$, and such a value which de Sitter took into account in 1929.}! In the solution A, however, there is \underline{no} radial velocity (...). We are thus confronted with the mathematical problem: what becomes of the empty world B if you fill it with matter. I have not yet been able to solve this problem completely but I have reasons to expect that the solution will be intermediate between the solution A and B \cite[AFA-FC-WdS-52]{de Sitter Archive}.
\end{quote}
Actually, such an intermediate solution, which interested de Sitter and, as we shall see, also Eddington, had been formulated by Lema\^{i}tre by 1927, but had passed unnoticed. Its important consequences were fully recognized by de Sitter and Eddington in 1930, when Lema\^{i}tre himself drew the attention of the two scientists to his own cosmological solution. In fact, in 1927 Lema\^{i}tre had proposed as a solution of relativistic field equations a non-empty, homogeneous, and isotropic universe where the spatial curvature radius increased in time. As shown by Lema\^{i}tre, it was the expansion of the universe which produced the cosmological redshift of galaxies.

\subsection{The debates on the empty universe and the ``de Sitter effect''}

Before 1930 it was the universe of de Sitter which stimulated debates and controversies, both for its puzzling properties, and for its actual relation to astronomical evidences. On the one hand, the static form of the de Sitter empty model showed a singularity: on the ``equator'' at the distance $r=\frac{\pi}{2}\,R$, as seen above, the potential term $g_{tt}$ was zero. Such a feature was criticized by Einstein, who did not accept the model formulated by de Sitter as a physical solution, and tried to discard it since it represented a counterexample of the relativity of inertia. The empty universe was a sort of an anti-Machian world model, where the inertia of a test particle was not determined by the world matter. Whereas de Sitter considered that such a surface was physically inaccessible to test particles, Einstein advocated the idea that it represented a ``mass-horizon''. According to Einstein, the universe of de Sitter was not really empty, but had matter concentrated at this equator. It was Felix Klein (1849-1925) who clarified this question, showing that this was a singularity due to the choice of coordinates, not a true physical singularity\footnote{We refer to \cite[pp. 351-357] {CPAE 1998} and \cite{Earman-Eisenstaedt 1999} for critical analysis of such discussions on the singularity in de Sitter's cosmological model. For the analysis of horizons in de Sitter's universe, see \cite{Rindler 1956}, for a description of horizons in modern cosmology, see \cite[chapter 21]{Harrison 2000}.}.

On the other hand, the de Sitter universe attracted the attention of scientists because, despite its lack of matter content, it offered more advantages than Einstein's one with respect to the astronomical consequences. In particular, de Sitter's static model offered an interpretation of the spectral displacement measured in distant objects by means of a twofold contribution. Namely, from de Sitter's analysis it emerged that such a cosmological solution predicted a spurious positive velocity of test particles due to the inertial field, superimposed to a real relative (Doppler) velocity. The latter contribution resulted from geodesic equations, and admitted both receding and approaching motions\footnote{By using the radial coordinate $\mathbf{r}$, the first contribution led to a quadratic redshift-distance relation, while the latter involved a linear dependence between spectral shift $z$ and distance $\mathbf{r}$ with no preference in sign: $z\simeq\frac{1}{2}\left(\frac{\mathbf{r}}{R}\right)^{2}\pm\frac{\mathbf{r}}{R}$. See \cite[pp. 195-196]{de Sitter 1933} for an analysis of such an effect given by de Sitter himself after the discovery of the expanding universe.}. It was such a twofold property, later known as the ``de Sitter effect'', which actually played during the Twenties the linking role between the relativistic description of the universe as a whole and the increasing amount of observational evidence of relevant radial velocities of spiral nebul{\ae}. As a matter of fact, in 1929 Hubble too referred to the possibility that the linear redshift-distance relation, revealed by his observations of galaxies, could actually represent the de Sitter effect, ``and hence that numerical data may be introduced into discussions of the general curvature of space'' \cite[p. 173]{Hubble 1929}.

The de Sitter universe can be regarded as the precursor of the non-static solutions to which attention was drawn after 1930. By quoting Merleau Ponty:
\begin{quote}
Statique sans l'\^{e}tre, vide mais non neutre, virtuellement actif sur toute
mati\`{e}re qu'on voudrait y mettre, r\'{e}sultat d'une sym\'{e}trie
en trompe-l'{\oe}il, solution b\^{a}tarde d'une \'{e}quation
b\^{a}tarde, l'univers de de Sitter \'{e}tait donc un curieux complexe d'\'{e}quivoques,
qui cependant portait l'avenir de la pense\'{e} cosmologiques \cite[p. 61]{Merleau Ponty 1965}.
\end{quote}
Since the geometry of de Sitter's universe is not uniquely specified\footnote{For clear descriptions of the many faces of the empty universe of de Sitter, see \cite{Schroedinger 1957,Ellis 1990}.}, during the Twenties many representations of this solution appeared, which complicated the actual interpretation of such a universe. Non-static versions of this empty model were formulated in 1922 by Kornel Lanczos (1893-1974) \cite{Lanczos 1922}, in 1923 by Weyl \cite{Weyl 1923a}, in 1925 by Lema\^{i}tre \cite{Lemaitre 1925}, and in 1928 by Howard Robertson (1903-1961) \cite{Robertson 1928}. In retrospect, these descriptions of de Sitter's universe correspond to truly expanding empty models, where the geometry of spatial sections at constant time is, respectively, positive ($k=+1$; Lanczos) and null ($k=0$; Weyl, Lema\^{i}tre, and Robertson) \cite[p. 100]{Ellis 1990}.

Each of these authors considered the theoretical redshift-distance relation, and looked for a proper formulation of the de Sitter effect. In particular, in his 1923 analysis of the hyperboloidal version of de Sitter's universe, Weyl obtained a relation which was roughly linear for small distances compared to $R$ \cite[p. 230]{Weyl 1923b}. Weyl proposed that in de Sitter's hyperboloid the world lines of test particles belonged to a pencil diverging from the past towards the future direction, which involved a general cosmic recession of nebul{\ae} in such a universe. This assumption on the choice of geodesics and their causal connection became later known as the ``Weyl principle''\footnote{We refer to \cite{Bergia-Mazzoni 1999,Ehlers 1988,Goenner 2001} for further readings on Weyl principle.}. A similar cosmic recession was suggested in the same year also by Eddington in his famous book \textit{The mathematical theory of relativity}, a compendium which later Einstein himself acknowledged as ``the finest presentation of the subject in any language'' \cite[p. 100]{Douglas 1956}. Eddington took into account the properties of the static form of de Sitter's universe, and found that a test particle could not remain at rest, but was accelerated away because of the presence of the cosmological constant. The preponderance of positive radial velocities of spirals, as revealed by some new observations of Slipher reported in Eddington's book, favored the model of de Sitter in comparison to Einstein's universe, which had matter but not motion. However, according to Eddington, these two rival solutions were ``two limiting cases, the circumstances of the actual world being intermediate between them. De Sitter's empty world is obviously intended as a limiting case; and the presence of stars and nebul{\ae} must modify it, if only slightly, in the direction of Einstein's solution'' \cite[p. 160]{Eddington 1923}.

An attempt to confirm such a cosmic recession was given by Carl Wirtz (1876-1939) with regard to the studies of the additional $K$ term found by Campbell for $B$ stars. Already in 1916 George Paddock (1879-1955) had investigated the determination of the direction of the solar motion by using now the radial motion of spiral nebul{\ae}, obtaining a $K$ term ranging from + 248 to + 295 km/sec \cite[p. 114]{Paddock 1916}. Wirtz further developed such an analysis on the velocities of spirals, and found the notable value of $K=+840\pm141$ km/sec \cite[p. 351]{Wirtz 1922}. In 1924, Wirtz introduced de Sitter's cosmology to account for such a large value of the additional $K$ term. He assumed that spirals had approximately the same linear diameter, so that their observed apparent diameter could be used as a distance indicator. Wirtz analyzed data of 42 spiral nebul{\ae}, for which both the apparent diameter and the radial velocity were known. He found a linear relation between the velocity and the logarithm of apparent diameter, which involved that the radial motion of spiral nebul{\ae} remarkably increased with increasing distance, as predicted by de Sitter's model \cite[pp. 23-24]{Wirtz 1924}.

However, in the same year the supposed general recession in the empty universe was strongly criticized by Silberstein, whose focus of interest was the determination of the curvature radius of de Sitter's space-time.

\section{Silberstein on globular clusters and de Sitter's universe}

Silberstein, a Polish-American physicist, maintained a sceptical approach towards many aspects of general relativity, which deserved him the role of one of the main critics of Einstein's theory\footnote{For some analysis of Silberstein's approach, see \cite{Desmet 2007,Flin-Duerbeck 2006,Sanchez Ron 1992}.}.

The determination of the curvature radius of de Sitter's universe was the subject of many papers written by Silberstein. Some considerations on the first two cosmological models were proposed by Silberstein already during the very early response to Einstein's new theory of gravitation. For instance, in the paper \textit{General relativity without the equivalence hypothesis}, which appeared in 1918, Silberstein acknowledged that the universe of de Sitter was particularly interesting because it did not involve any hypothetical world matter, which on the contrary was unavoidable in Einstein's model \cite[p. 105]{Silberstein 1918}. In this paper Silberstein remarked that the general covariance was one of the very strong points of Einstein's theory of gravitation. The cosmological solution proposed by de Sitter was therefore preferable to Einstein's universe, since it fully achieved this requirement, and was perfectly homogeneous and isotropic, as explicitly stated by Silberstein some years later \cite[p. 67]{Silberstein 1930}.

The objection to the world matter was reported by Silberstein also in his 1922 book on the theory of general relativity and gravitation. Here Silberstein noticed that, for a density of matter of the order of ``some thousand suns per cubic parsec'', the curvature radius of Einstein's universe could not be smaller than $10^{12}$ AU, and consequently one should admit the existence of $10^{10}$ galaxies filling this space. On the contrary, the possibility to interpret spectral shifts as predicted by de Sitter represented an ``attractive piece of reasoning'' \cite[p. 137]{Silberstein 1922}.

In order to determine the size of de Sitter's universe, Silberstein used in 1924 an approximately linear relation between spectral shift and distance.
However, he polemically criticized the general tendency of
particles to scatter in de Sitter's universe, and proposed a
theoretical relation which was valid for both receding and approaching objects.
According to Silberstein, the recession in de Sitter's universe formulated by Weyl in 1923 was
``an arbitrary hypothesis'' \cite[p. 909]{Silberstein 1924c}; furthermore,
the ``mythical'' assumption \cite[p. 350]{Silberstein 1924a}
that the world lines belonged to a pencil of geodesics diverging
towards the future was a ``sublime guess, entirely undesirable'' \cite[p. 909]{Silberstein
1924c}. Eddington's suggestion on a
universal scattering of test particles was also considered by Silberstein
``a fallacy based upon a hasty analysis''  \cite[p. 350]{Silberstein
1924a}.

In fact, the recession advocated by Eddington and Weyl was
contradicted by the negative velocities of some spirals.
Among them, the blueshift measured in the spectrum of the Andromeda
nebula revealed a relevant approaching motion with a velocity of the order of $v\simeq\,-316$ km/sec.
Therefore, Silberstein based his analysis on the observations of globular clusters.
Actually, such objects showed both receding and approaching motions equally distributed. Furthermore, Silberstein
acknowledged that the estimates of the radial velocity of globular clusters were known with small probable error in comparison to spirals. Moreover, despite the attempts made, among others, by Lundmark in 1919 \cite{Lundmark 1919} and by Ernst \"{O}pik (1893-1985) in 1922 \cite{Opik 1922} to determine the distance of the Andromeda nebula, there was not (yet) a general consensus on reliable estimates of the distance of spirals.

In his papers on the size of the universe, Silberstein referred to the works by Shapley on the observations of globular clusters and on the estimate of the size of the Milky Way. In this respect, the historical reconstruction proposed in \cite[p. 144]{Smith 1979} is useful to reveal the role played by Shapley, who initially encouraged Silberstein to investigate the de Sitter effect, but later showed less interest in his results.

It is worth recalling that, some years before, Shapley had given a fundamental contribution to the comprehension of the structure of the Milky Way. Shapley used statistical parallaxes in order to determine the absolute magnitude of some RR Lyr{\ae} stars, i.e. pulsating
variable stars, like Cepheids, which change in brightness with a
regular period. Shapley was able to estimate the distance of
these stars observed in globular clusters by means of the period-luminosity
relation discovered in 1912 by Hernrietta Leavitt (1868-1921). In
1919, Shapley set the diameter of our Galaxy of about 300,000 light-years ($19\cdot10^{9}$ AU). The center of the Galaxy, according to Shapley, was 65,000 light-years ($4\cdot10^{9}$ AU) far
from the Sun, in the direction of Sagittarius\footnote{The main features of our Galaxy have not changed much since those proposed in 1936 by John Plaskett (1865-1941). According to Plaskett, the Milky Way is a flat rotating disk, with a diameter of about 100,000 light years ($6\cdot10^{9}$ AU), surrounded by a spherical halo of globular clusters \cite{Plaskett 1936}.}. Furthermore, Shapley furnished several topics against the extragalactic interpretation of spiral nebul{\ae}, which he proposed in 1920 during the so-called ``Great Debate'', the famous discussion between himself and Heber Curtis (1872-1942), focused on the size of the Milky Way and the nature of spiral nebul{\ae}\footnote{We refer to \cite{Hoskin 1976} for a historical reconstruction of the ``Great Debate''.}. On the one hand, Curtis advocated the theory that spirals were truly external galaxies \cite{Curtis 1920}. On the other hand, Shapley remarked that ``we have no evidence that somewhere in space there are not other galaxies; we can only conclude that the most distant sidereal organizations now recognized (globular clusters, Magellanic Clouds, spiral nebul{\ae}) cannot successfully maintain their claims to galactic structure and dimensions'' \cite[p. 268]{Shapley 1919}\footnote{It is worth noting that Silberstein did not agree with Shapley's conclusion: ``it would certainly be foolish - Silberstein wrote in 1922 - to deny the possibility (...) of the existence of many more island universes'' \cite[p. 134]{Silberstein 1922}.}.

It is worth mentioning that the determination of the distance of Cepheid stars represents an important step in the cosmological distance ladder, the construction of which is fundamental in astronomy as well as in cosmology. In the present picture, the cosmic ladder is made of distinct steps, obtained by using different methods: from trigonometric parallax and kinematic methods for distances within the Galaxy, to primary and secondary indicators for extragalactic distances, as, for instance, RR Lyr{\ae} and Cepheid variable stars, Nov{\ae}, Supergiants, Supernov{\ae}, globular clusters, $HII$ regions (i.e. clouds of ionized hydrogen), brightest cluster galaxies\footnote{For further readings on the cosmological distance ladder, see \cite{Webb 1999}. For a review of the methods to determine extragalactic distances, see \cite{Freedman-Madore 2010}.}. In the investigation of astronomical distances, some steps were identified after the discovery of the expanding universe. In 1930, for instance, Robert Trumpler (1886-1956) unambiguously confirmed the existence of the interstellar absorption of light affecting astronomical observations \cite{Trumpler 1930}. In 1934, Walter Baade (1893-1960) and Fritz Zwicky (1898-1974) suggested the use of Supernov{\ae} as potential distance indicators \cite{Baade-Zwicky 1934}. In 1952, Baade provided evidence for a new calibration of the extragalactic distance scale, based on his discovery in 1944 of the existence of two stellar populations, presenting two types of Cepheid variable stars \cite{Baade 1944,Baade 1952}\footnote{For a description of the early history of the period-luminosity relation, see \cite{Fernie 1969}. For Baade's contributions to astrophysics, see \cite{Osterbrock 2001}.}.

Silberstein showed great interest in the possibility to obtain further empirical observations. This aspect is revealed for instance by the 1924 correspondence between himself and Walter Adams (1876-1956), one of the member of the American section of the Committee on stellar radial velocities, whom Silberstein (unsuccessfully) asked for obtaining velocities of 74 globular clusters. In June 1924, Silberstein wrote to Adams that:
\begin{quote}
The knowledge of radial velocities of remote objects of ascertainable distance became an urgent need, and the (spectroscopic) measurement of these velocities should, in my opinion, be incorporated into the programme of your Committee. Actually, since the spiral nebul{\ae} baffle all attempts at estimating their distance, the objects in question are the globular clusters \cite[62.108]{Adams Archive}.
\end{quote}
Silberstein considered the line element of the static form of de
Sitter's universe. Despite his critical remarks to Weyl's conclusion on the alleged cosmic recession, Silberstein used the same general
principle formulated in 1923 by Weyl himself, who related the spectral shift $z$ to the ratio of the proper time $ds$ of the observer to the proper time $ds'$ of the source: $z=\frac{ds}{ds'}-1$. Silberstein obtained for what he called ``the complete Doppler effect'', i.e. the de Sitter effect, the relation:
\begin{equation}
z=\gamma\left[1\pm\sqrt{1-\frac{\cos^{2}\frac{r}{R}}{\gamma^{2}}}\right]-1,
\end{equation}
where $\gamma=\left(1-\frac{v^{2}}{c^{2}}\right)^{-1/2}$. The positive sign corresponded to receding objects, while the
negative sign to approaching ones \cite[p. 912]{Silberstein 1924c}.

In such a general formula of the Doppler effect there were
two terms: a term depending on the velocity $v$,
which was dominant near the observer, and a second term depending
upon $\frac{r}{R}$, which, according to Silberstein, was significant
for very remote celestial objects. For near stars, the velocity
effect approximated to the special relativistic one.
On the contrary, for the most distant celestial objects, the relation became \cite[p. 351]{Silberstein 1924a}:
\begin{equation}
z\simeq\pm\frac{r}{R}.
\end{equation}
Silberstein used such a linear relation in order to
determine the value of the curvature radius of de Sitter's world. He took into account
radial velocities, both positive and negative, and distances of
seven globular clusters, and obtained a mean value of
$R=6\cdot10^{12}$ AU \cite[p. 351]{Silberstein 1924a}. Such a result was almost confirmed by using velocity
and distance of the two Magellanic Clouds \cite[p. 363]{Silberstein
1924b}.

The most distant spiral which was known at that time, NGC 584,
showed a radial velocity of about $v=+1800$ km/sec.
Therefore it followed that such an object was placed at the distance
$r=3.6\cdot10^{10}$ AU. ``Huge as this may seen -
Silberstein noted - it will be remembered that Shapley's latest
estimate of the semi-diameter of our galaxy is only four times
smaller. (...) Whether these estimates will or will not fit into the
general scheme of modern galactic and extra-galactic astronomy, is
not known to me and must be left to the scrutiny of specialists''
\cite[pp. 916-917]{Silberstein 1924c}.

Later on, Silberstein showed that now, from the velocity and
distance of ten objects, i.e. eight clusters and the Magellanic Clouds,
a linear relation was actually confirmed by plotting, as suggested to him
by Henry N. Russell (1877-1957), the modulus of the redshift: $r=|z|\,R$.
Silberstein, however, discarded data belonging to other three globular clusters
(NGC 5904, NGC 6626, NGC 7089), whose velocities were ``suspiciously small' and did not give a constant curvature radius
\cite[p. 602]{Silberstein 1924d}. From data of these ten objects,
the size of the universe, by using the general Doppler
formula, was of the order of $R\geq\,9.1\cdot10^{12}$ AU, while the approximate linear
formula led to a world radius of de Sitter's universe not exceeding
$R=8\cdot10^{12}$ AU \cite[p. 819]{Silberstein 1924e}. Silberstein also employed and further elaborated
a statistical formula in order to get $R$ in terms of the mean $z$ and $r$ of two groups of objects for which the mean velocity was the same:
\begin{equation}
\bar{z}_{2}^{2}-\bar{z}_{1}^{2}=\frac{2}{3R^{2}}(\bar{r}_{2}^{2}-\bar{r}_{1}^{2}).
\end{equation}
The bars denote average values, and the suffixes the different groups. By splitting thirteen objects in two groups of, respectively, seven and six objects, such an analysis gave
$R=7.2\cdot10^{12}$ AU \cite[p. 627]{Silberstein 1924f}. Such a method was criticized by Eddington, who pointed out that the derivation of such a formula disagreed with Lorentz transformation \cite[p. 747]{Eddington 1924}.

As we shall see in the next section, the Swedish astronomer Lundmark disapproved of Silberstein's analysis, which he found objectionable both for the choice of a \textit{selected} number of globular clusters, and for the supposed linear correlation between shift and distance. Already in 1924, Lundmark proved that the methods and the results by Silberstein were wrong, so that they did not much appeal to scientists involved in the early debates on relativistic cosmology. Nevertheless, the effort made by Silberstein stimulated further investigations on de Sitter's model, as revealed for instance by the 1925 work on this subject by Lema\^{i}tre, and later by the contributions of Robertson and Richard Tolman (1881-1948), which appeared, respectively, in 1928 and 1929. In this respect, it is worth to quote part of the draft of an obituary for Robertson written by Lema\^{i}tre in 1963. Here Lema\^{i}tre returned to his 1925 interpretation of the non-static feature of de Sitter's universe, and acknowledged that:
\begin{quote}
I was better prepared to accept it following an opinion expressed by Eddington. (...) The
errors by Silberstein have been very stimulating. I had myself had a
long discussion with him in 1924 at a British
Association Conference in Toronto and my work, as possibly later on
the work of Robertson, results as a large part as a reaction against
some unsound aspects of Silberstein's theories \cite[D32]{Lemaitre Archive}.
\end{quote}
The interest of Silberstein in this topic culminated in his book \textit{The size of the universe}, which was written in 1929 and published in 1930, i.e. just at the turning point of the discovery of the expanding universe \cite{Silberstein 1930}. Here Silberstein collected his considerations on relativistic cosmology. He maintained the objection to the general tendency of particles to scatter suggested by Weyl, and criticized some measurements made by Hubble of the distance of extragalactic nebul{\ae}. Silberstein pursued his analysis on the static metric of de Sitter's universe, and kept accepting the proposal of a theoretical relation valid for both red and blue shifts in order to obtain a constant curvature radius (see table \ref{final} for further estimates of $R$ reported by Silberstein in his book). Eddington sharply stated that the views held by Silberstein on a finite and static universe were obsolete, being now superseded by the much more interesting proposal on the expanding universe made by Lema\^{i}tre \cite[p. 850]{Eddington 1930}. Silberstein's book was later criticized also by Robertson. According to Robertson, Silberstein had not been able to account for the ``overwhelming preponderance of redshifts'' revealed by the works of Hubble and Humason \cite[p. 603]{Robertson 1932}.

\section{The analysis proposed by Lundmark}

In August 1924 Lundmark wrote a paper on the determination of the curvature radius of de Sitter's space-time. The detailed analysis given by Lundmark is a clear example of the empirical approach adopted by the Swedish astronomer, based on the accurate review of available data, on the systematic comparison of different independent observations, and on the prompt use of working hypotheses.

Lundmark started his analysis by carefully examining the question of the nature of measured shifts, and proved that the spectral displacement was nearly constant for 16 lines in the Andromeda nebula. Such a shift, according to Lundmark, was thus a Doppler one, as well as the shift measured in globular clusters. However, he claimed that the origin of such displacements was still uncertain.

Lundmark was sceptical on the alleged possibility that the motion of globular clusters showed any effect of the curvature of space-time, and criticized the method proposed by Silberstein, who ``has not given, and will probably not be able to give, any justification for the use of the velocities of the globular clusters for a determination of $R$'' \cite[p. 750]{Lundmark 1924}. A small $K$ term resulted from the analysis of 18 globular clusters, while a larger value was obtained by treating velocities of 43 spiral nebul{\ae}. Therefore Lundmark asserted that globulars were nearer than spirals. As a consequence, spirals could presumably be affected by the curvature of space-time, whereas the motion of globular clusters was a real phenomenon, and could not be interpreted as the spurious velocity predicted for distant objects by the de Sitter effect. Moreover, Silberstein's result was objectionable because of the selected choice of those radial velocities of globular clusters which gave a constant value of the curvature radius. Lundmark made use of his own observations of 18 globulars, and compared his data to Shapley's ones. He claimed that his own analysis superseded the one presented by Silberstein, and concluded that there was not any definite correlation between velocity and distance. In addition, by hypothetically admitting the validity of Silberstein's linear relation, the mean value of the curvature radius resulted to be $R\simeq\,19.7\cdot10^{12}$ AU, nearly three times larger than the radius calculated by Silberstein, and in any case a still larger radius was likely to be expected\footnote{As later noted by Silberstein, in such a 1924 paper Lundmark erroneously reported the unit of distance in km rather than in AU \cite[p. 285]{Silberstein 1925}.}.

In the second part of his paper Lundmark dealt with further possibilities to determine the radius $R$ by means of several classes of stars, which he used as distance indicators. He reviewed data of velocity and distance belonging to 30 Cepheid stars, 8 Nov{\ae}, 27 $O$ stars, 29 $R$ stars, 25 $N$ stars, and 31 Eclipsing variables. With regard to the distance of Cepheid stars, Lundmark followed the derivation proposed by Shapley, who studied their proper motion together with the period-luminosity law. Distances of Nov{\ae} were determined by assuming that the mean absolute maximum magnitude had almost a constant value. When the radial velocities were plotted according to the corresponding distances, there seemed to be no progression, while, on the contrary, one should expect a progression from Silberstein's analysis on the constant curvature radius. The average values of the curvature radius found by Lundmark (by applying Silberstein's formula to the different classes of objects mentioned above) were, respectively, 7.5, 41, 4.0, 6.7, 2.3, $2.7 \cdot10^{12}$ AU \cite[pp. 756-763]{Lundmark 1924}.

The role played by spiral nebul{\ae} as distance indicators was the subject of the last part of Lundmark's paper, where the Swedish scientist proposed a pioneering empirical analysis of the relation between the velocity and distance for 44 spiral nebul{\ae}. Already in 1919, as mentioned before, Lundmark had estimated the distance of the Andromeda nebula at 200,000 pc ($4\cdot10^{10}$ AU) by means of the Nov{\ae} maximum brightness method. Now he used such a value as the unit of the distance scale. The parallax of spirals was obtained by using the working hypothesis ``that the apparent angular dimensions and the total magnitudes of the spiral nebul{\ae} are only dependent on the distance'' \cite[p. 767]{Lundmark 1924}. Lundmark also applied the statistical method, developed by Silberstein, to two groups of, respectively, 23 and 18 objects, which gave $R=2.4 \cdot10^{12}$ AU \cite[p. 769]{Lundmark 1924}. However, the conclusion reached by Lundmark was that the values of the curvature radius derived from each single spiral were exceedingly different, and thus inconsistent. Nevertheless, he found that there seemed to be a relation between radial velocity and distance, ``although not a very definite one'' \cite[p. 768]{Lundmark 1924}.

In a subsequent paper, published in 1925, Lundmark offered a complete review of the direct and the indirect methods to estimate the distance of spiral nebul{\ae}. It was at the end of this paper that the Swedish astronomer returned to question of the extension of the universe. Here he analyzed some of his observations in the light of the static, infinite and hierarchical model of the universe supported in those years by Charlier, another Swedish astronomer. Recalling some ideas by Johann Heinrich Lambert (1728-1777), Charlier proposed that celestial bodies formed gradually increasing spherical galaxies:
\begin{itemize}
\item $N_{1}$ stars formed galaxy $G_{1}$, of order 1 and radius $R_{1}$
\item $N_{2}$ galaxies $G_{1}$ formed galaxy $G_{2}$,
of order 2 and radius $R_{2}$
\end{itemize}
and so forth... \cite[p. 186]{Charlier 1925}.
By means of counts of stars and nebul{\ae}, and of the
apparent dimension of the latter, Charlier obtained the relation:
\begin{equation}
\frac{R_{i}}{R_{i-1}}>\sqrt{N_{i}}.
\end{equation}
A nearly altered version of such a relation was useful to estimate the distance of the Andromeda nebula (NGC 224), since, according to Charlier, spirals were galaxies of the second order \cite[p. 892]{Lundmark 1925}:
\begin{equation}
\frac{R_{2}}{R_{1}}=\sqrt{N_{2}}.
\end{equation}
Lundmark found a rough agreement between Charlier's result on the Andromeda distance (28 times the diameter of the galactic system), and his own result (32 times the galactic diameter). ``Our present knowledge - Lundmark thus emphasized - as to the space-distribution of the stars and the spirals can be summed up in the statement: \textit{our stellar system and the system of spiral nebul{\ae} are constructed according to the conceptions expressed in the Lambert-Charlier cosmogony}'' \cite[p. 893]{Lundmark 1925}.

\section{Further determinations of the curvature radius before the expanding universe}

As seen in the previous sections, at the beginning of relativistic cosmology scientists as de Sitter, Silberstein, and Lundmark showed great interest in the determination of $R$. In addition to their systematic analysis in terms of astronomical observations, further considerations on the size of the universe appeared in different frameworks. For instance, Weyl and Eddington took into account the curvature radius of space-time in their speculative attempts to connect macro-systems with micro-systems, in particular in the perspective of large numbers coincidence\footnote{On the large numbers hypothesis, see for instance \cite{Barrow 1990}. We refer to \cite{Gorelik 2002} for an analysis of Weyl's considerations on large numbers in relativistic cosmology.}. In fact, Weyl considered the relation between the world radius, the radius of the electron, and the gravitational radius associated with a mass $m$. Eddington, who already in August 1917 had emphasized in a letter to de Sitter that ``it is very interesting that you can get a determination of the necessary order of magnitude of $R$'' \cite[AFA-FC-WdS-11]{de Sitter Archive}, followed Weyl in this analysis. In 1920, right in the light of Weyl's approach on the unification of electricity and gravitation, Eddington argued that $R$ of Einstein's world was of the order of $2\cdot10^{11}$ pc ($4\cdot10^{16}$ AU), ``which - Eddington noted - though somewhat larger than the provisional estimates made by de Sitter, is within the realm of possibility'' \cite[p. 179]{Eddington 1920}. Later on, a numerical value of $R$ was reported by Weyl in the appendix of the fifth edition of his \textit{Raum, Zeit, Materie}. Here Weyl, in the already mentioned investigation of the hyperboloidal version of de Sitter's universe, referred to Lundmark's result on the distance of the Andromeda nebula, and found that $R=10^{9}$ AU: the curvature radius was $10^{40}$ times the radius of the electron, which was the same ratio of this latter to the gravitational radius of the electron \cite[p. 323]{Weyl 1923a}.

It is worth mentioning Eddington's further attempts to relate the cosmological problem to the atomic one. Eddington considered the cosmological constant as one of the fundamental entities in nature, together with the fine structure constant, the number of particles expected in an expanding universe, and the ratio of electrostatic and gravitational forces. For instance, in 1931 Eddington suggested that an estimate of $\lambda$ could be obtained by means of the wave equation for an electron, in which both the number of electrons in the universe and the time-dependent world radius should appear \cite{Eddington 1931}. Some years later, Paul Dirac (1902-1984) too followed the approach of investigating in the cosmological framework the mathematical relations which, according to him, connected the large dimensionless numbers occurring in nature \cite{Dirac 1938}.

As a matter of fact, the 1924 authoritative contribution proposed by Lundmark on the determination of $R$ revealed that a common criterion for the choice of distance indicators was missing. The failure of Silberstein's analysis influenced the cosmological debate of the late twenties, in the sense that the interest shown in the size of the universe gradually decreased, while the attention of modern cosmologists was mainly focused on the nature of spiral nebul{\ae} and their role in testing cosmological models. Gustaf Str\"{o}mberg (1882-1962) gave in 1925 a comprehensive analysis of the velocity of globular clusters and ``non-galactic nebul{\ae}''. In fact, Str\"{o}mberg confirmed that the interpretation of the relevant redshifts and the form of the redshift-distance relation represented an issue still to be clarified \cite{Stromberg 1925}. In a 1925 summary of the different attempts to measure the size of the universe, Archibald Henderson (1877-1963) concluded that:
\begin{quote}
If, as now appears probable, the spirals are isolated systems, this recession must be explained, it appears, either as a wholesale error or else as a relativistic effect (...). Much additional data will be required and many further researches made before it will be possible categorically to decide between the infinite, limitless, Euclidean universe of Newton, and the finite, unbounded, non-Euclidean universe of Einstein and de Sitter \cite[p. 223]{Henderson 1925}.
\end{quote}
In this framework, the contributions proposed by Hubble marked a second renewal of cosmology. On the one hand, the determination in 1925 of the distance of the Andromeda nebula by means of Cepheid variables disclosed the depth of the realm of the galaxies \cite{Hubble 1925}\footnote{As noted above, the measurements of the distance of galaxies made by Hubble were reconsidered after Baade's discovery of the existence of two different stellar populations.}. On the other hand, the linear redshift-distance relation, formulated in 1929, held the evidence of a nearly \textit{systematic} recession of distant galaxies \cite{Hubble 1929}.

Actually, between the years 1925-1930, only few suggestions on the dimension of the universe appeared in scientific papers. Among them, Hubble proposed a value of $R$ of Einstein's universe in the last section of his 1926 paper devoted to the general classification of extragalactic nebul{\ae}. Here Hubble calculated that the mean density of world matter was $\rho=1.5\cdot10^{-31}$ g/cm$^{3}$, which involved that $R=2.7\cdot10^{10}$ pc ($5.6\cdot10^{15}$ AU) \cite[p. 369]{Hubble 1926}.

An estimate of $R$, as noted in \cite[p. 151]{Kragh 2007}, was found in the same year by Wilhelm Lenz (1888-1957), now in relation to thermodynamics equilibrium. Lenz applied to the volume of the Einstein universe the 1925-26 analysis proposed by Otto Stern (1888-1969) on the relation between the energy density of matter, the energy density of black body radiation, and the temperature of space. The result, which Lenz acknowledged to be of a ``fascinating simplicity'', was that, at the equilibrium, the matter energy was equal to the radiation energy. The relation proposed by Lenz between the temperature $T$ and $R$ was:
\begin{equation}
T^{2}=\frac{1}{R}\left(\frac{2c^{2}}{a\kappa}\right)^{1/2}\simeq\frac{10^{31}}{R},
\end{equation}
where $a$ is the Stefan-Boltzmann constant, and $R$ is expressed in cm. As Lenz noted, the density of world matter of the order of $10^{-26}$ g/cm$^{3}$ led to $R=10^{26}$ cm ($6.7\cdot10^{12}$ AU), and consequently the black body temperature was too high, about $T = 300 K$. On the contrary, by assuming in such a relation the radiation temperature of 1 $K$, $R$ resulted to be about $10^{31}$ cm ($6.7\cdot10^{17}$ AU) \cite[p. 644]{Lenz 1926}.

Two years later, Robertson considered the distance of spirals reported in Hubble's 1926 paper, in relation to Slipher's data of radial velocity reported in Eddington's 1923 book on relativity. According to Robertson, these observations were able to confirm a nearly linear redshift-distance relation which Robertson had derived by his own non-static version of the metric of de Sitter's universe. By means of such data, Robertson found the curvature radius of the empty universe to be $R=2\cdot10^{27}$ cm ($1.3\cdot10^{14}$ AU) \cite[p. 845]{Robertson 1928}.

In 1929, Tolman offered a very detailed analysis of the static form of de Sitter's line element, in which he investigated the properties of a formula of the de Sitter effect more general than the one previously found by Silberstein. Tolman assumed that $R$ would be of, at least, ten times the range of the most distant galaxies: $R\geq2\cdot10^{8}$ light-years ($1.2\cdot10^{13}$ AU) \cite[p. 271]{Tolman 1929}. Furthermore, in order to reconcile the general tendency of scatter in de Sitter's universe, Tolman introduced the hypothesis of continuous entry (even continuous formation), namely that ``nebul{\ae} are continually entering, as well as leaving the range of observation'', from which he concluded that $R=2\cdot10^{9}$ light-years ($1.2\cdot10^{14}$ AU) \cite[p. 272]{Tolman 1929}.

A further estimate of $R$ was proposed yet in 1927, but in a different context, i.e. within the considerations on the time-dependent world radius which appeared in the paper on the expanding universe written in that year by Lema\^{i}tre. As a matter of fact, it was Friedmann who first showed in 1922 that Einstein's and de Sitter's solutions were ``special cases of more general assumptions'', and then demonstrated ``the possibility of a world in which the curvature of space is independent of the three spatial coordinates but does depend on time'' \cite[p. 49]{Friedmann 1922}. However, Friedmann did not relate his theoretical predictions to astronomical observations. On the contrary, this was done in 1927 by Lema\^{i}tre, who took into account empirical data in order to calculate $R(t)$. In Lema\^{i}tre's paper, a partial English translation of which was diffused in 1931, the Belgian scientist had pointed out that ``in order to find a solution combining the advantages of those of Einstein and de Sitter, we are led to consider an Einstein universe where the radius of space or of the universe is allowed to vary in an arbitrary way'' \cite[p. 484]{Lemaitre 1931}. The radius $R\equiv\,R(t)$ asymptotically increased with time, starting from $R_{0}=\lambda^{-1/2}$, which value depended on the cosmological constant and was the radius at $t=-\infty$ \cite[p. 94]{Lemaitre 1927}. In fact, it was Lema\^{i}tre who offered in his 1927 paper a solution to the puzzling interpretation of relevant redshifts. He clearly stated that such spectral displacements in spirals were a cosmical effect due to the variation of $R$, i.e. to the expansion of the universe. The redshift-distance relation valid for near objects was \cite[p. 96]{Lemaitre 1927}:
\begin{equation}
z=\frac{v}{c}\simeq\frac{R_{2}-R_{1}}{R_{1}}=\frac{R'}{R}r,
\end{equation}
where $R_{1}$ and $R_{2}$ were, respectively, the radius of the universe at the time of emission of a light signal, and that at the epoch of reception, and $R'$ referred to the derivative of $R$ with respect to time. Lema\^{i}tre calculated the distance of 43 nebul{\ae} by assuming, as done by Hubble in 1926, that they had the same absolute magnitude. The average distance he found was $r=10^{6}$ pc ($2\cdot10^{11}$ AU) \cite[p. 96]{Lemaitre 1927}. With regard to the radial velocity, the Belgian scientist referred to the observations collected in 1925 by Str\"{o}mberg and in 1926 by Hubble, and assumed  $v$ = 625 km/sec as the average velocity at this distance\footnote{Actually, this can be seen as the first suggestion of the value of what became later known as the ``Hubble constant'', in the sense that here Lema\^{i}tre stated that, by assuming the proportionality between $v$ and $r$, a galaxy observed at the distance of 1 Mpc would recede with a velocity of 625 km/sec. However, the section containing these values and calculation proposed by Lema\^{i}tre in 1927 was not reported in the 1931 English translation. In 1929, Hubble, unaware of Lema\^{i}tre's result, obtained for this constant, i.e. for $K$ term in his linear relation $v=Kr$, a value ranging from +465 to +530 km/sec \cite[pp. 170-172]{Hubble 1929}. We refer to \cite{Trimble 1996} for a historical reconstruction of the determination of the Hubble constant from 1925 to 1975.}, whether the range was between +575 and +670 km/sec \cite[p. 97]{Lemaitre 1927}. In this way Lema\^{i}tre found:
\begin{equation}
\frac{R'}{R}\equiv\,y=0.68\cdot10^{-27}\,\textrm{cm}^{-1},
\end{equation}
from which it followed:
\begin{equation}
R=R_{A}\sqrt{y}=6\cdot10^{9}\,\textrm{pc}\,(\simeq1.2\cdot10^{15}\,\textrm{AU}),
\end{equation}
\begin{equation}
R_{0}=R_{A}\,y^{3/2}=2.7\cdot10^{8}\,\textrm{pc}\,(\simeq5.6\cdot10^{13}\,\textrm{AU}).
\end{equation}
Here $R_{A}$ was the constant radius of the Einstein universe, for which Lema\^{i}tre used the value determined by Hubble in 1926 ($R_{A}=2.7\cdot10^{10}$ pc\,$\simeq5.6\cdot10^{15}$ AU) \cite[p. 98]{Lemaitre 1927}.

The relativistic field equations in the form derived by Friedmann and independently by Lema\^{i}tre, known as the ``Friedmann-Lema\^{i}tre equations'', related the time-dependent world radius to the world matter content and the cosmological constant, and were able to describe the evolution of the expanding universe (later on, in such equations $R(t)$ was substituted by $a(t)$, which refers to the expansion parameter, or cosmic scale factor). Finally, in view of the cosmic recession of galaxies from each other, empirically confirmed by Hubble's observations, the rediscovery in 1930 of the dynamical models of the expanding universe formulated by Friedmann and Lema\^{i}tre inaugurated a new phase in the modern understanding of the universe as a whole.

\begin{table}

\caption{Summary of the estimates of the size of the universe from 1917 to 1930 ($R_{A}$ and $R_{B}$ refer to, respectively, the model of the universe of Einstein, and that of de Sitter). One of the currently accepted estimates of the curvature scale $R_{c}$ of the universe is $R_{c}>42$ Gpc\, ($\simeq87\cdot10^{14}$ AU). It is worth noting that $R_{c}$ is now expressed as a function of the Hubble constant and the energy-matter density of the universe (for details, see \cite{Vardanyan-Trotta-Silk 2011}).}\label{final}

\begin{center}
\begin{tabular}{llllr}
\hline
author & year & method or astronomical objects & radius & AU\\
\hline
 Einstein    & 1917   & matter  density  (stars)     & $R_{A}=$ & $6\cdot10^{11}$\\
 de Sitter   & 1917   & galaxy apparent diameter     & $R_{A}\geq$ & $10^{12}$\\
             &        & matter density (stars)                 & $R_{A}=$ & $9\cdot10^{11}$\\
             &        & matter  density (galaxies)   & $R_{A}\leq$ & $5\cdot10^{13}$\\
             &        & light absorption         & $R_{A}>$ & $ 1/4\cdot10^{12}$\\
             &        & $B$ stars                    & $R_{B}=$ & $2/3\cdot10^{10}$\\
             &        & Small Magellanic Cloud       & $R_{B}>$ & $2\cdot10^{11}$\\
             &        & 3 galaxies                   & $R_{B}=$ & $3\cdot10^{11}$\\
 Eddington   & 1920   & large numbers hypothesis     & $R_{A}=$ & $4\cdot10^{16}$\\
 Silberstein & 1922   & matter density (galaxies)    & $R_{A}\geq$ & $10^{12}$\\
 Weyl        & 1923   & Andromeda galaxy             & $R_{B}=$ & $10^{9}$\\
 Silberstein & 1924   & 7 globular clusters          & $R_{B}=$& $6\cdot10^{12}$\\
             &        & 8 globular clusters + Magellanic Clouds & $R_{B}\geq$ & $9.1\cdot10^{12}$\\
             &        & 11 globular clusters + Magellanic Clouds& $R_{B}=$ & $7.2\cdot10^{12}$\\
 Lundmark    & 1924   & 18 globular clusters         & $R_{B}=$ & $19.7\cdot10^{12}$\\
             &        & Cepheid stars                & $R_{B}=$ & $7.5\cdot10^{12}$\\
             &        & Nov{\ae} stars               & $R_{B}=$ & $41\cdot10^{12}$\\
             &        & $O$ stars                    & $R_{B}=$ & $4\cdot10^{12}$\\
             &        & $R$ stars                    & $R_{B}=$ & $6.7\cdot10^{12}$\\
             &        & $N$ stars                    & $R_{B}=$ & $2.3\cdot10^{12}$\\
             &        & Eclipsing variable stars     & $R_{B}=$ & $2.7\cdot10^{12}$\\
             &        & 41 galaxies                  & $R_{B}=$ & $2.4\cdot10^{12}$\\
 Hubble      & 1926   & matter density (galaxies)    & $R_{A}=$ & $5.6\cdot10^{15}$\\
 Lenz        & 1926   & radiation temperature        & $R_{A}=$ & $6.7\cdot10^{17}$\\
 Lema\^{i}tre& 1927   & 43 galaxies                  & $R(t)$= & $1.2\cdot10^{15}$\\
 Robertson   & 1928   & 42 galaxies                  & $R_{B}=$ & $1.3\cdot10^{14}$\\
 Tolman      & 1929   & galaxies                     & $R_{B}\geq$ & $1.2\cdot10^{13}$\\
             &        & continuous entry hypothesis  & $R_{B}=$ & $1.2\cdot10^{14}$\\
 de Sitter   & 1929   & matter density (galaxies)    & $R_{A}=$ & $3\cdot10^{13}$\\
 Silberstein & 1930   & 18 globular clusters + Magellanic Clouds & $R_{B}=$ & $7.4\cdot10^{12}$\\
             &        & 29 Cepheid stars             & $R_{B}=$ & $3\cdot10^{11}$\\
             &        & 35 $O$ stars                 & $R_{B}=$ & $3.2\cdot10^{11}$\\
             &        & 459 stars                    & $R_{B}=$ & $4\cdot10^{11}$\\
\hline
\end{tabular}
\end{center}
\end{table}

\section{Conclusion}

The period ranging from the 1917 static cosmological model of Einstein to the 1930 official entrance of the proposal of the expanding universe was characterized by a variety of ideas, discoveries, and controversies.

In those years new perspectives were opened in the far-reaching and still ongoing challenge to the comprehension of the universe by means of the laws of physics. The leading scientists dealt with several issues which emerged from the first tortuous, but at the same time fruitful, interplay between predictions of general relativity and astronomical observations. It is noteworthy that some topics faced in those years are still present in the cosmological debates. We mention, for instance, issues as the postulate of the homogeneity and isotropy of space, the existence of visual horizons, as well as the cosmological interpretation of spectral shifts, and the use of distance indicators for extragalactic objects.

Certainly, the interest in the determination of the size of the universe faded away as the expanding universe entered modern cosmology. Nonetheless, as shown in the present analysis, the contributions given from 1917 to 1930 to this specific question can be seen as an interesting example of the efforts that were made to achieve a coherent picture of the universe by understanding, in agreement with the legacy of Galileo, its \textit{mathematical language} in the light of the \textit{sensible experiences}.

``The theory of today - de Sitter wrote in 1932 - is not the theory of tomorrow. (...) Science is developing so very rapidly nowadays, that it would be preposterous to think that we had reached a final state in any subject. The whole of physical science, including astronomy, is in a state of transition and rapid evolution. Theories are continually being improved and adapted to new observed facts. It would certainly not be right to suppose at the present time that we had reached any state of finality. We are, however, certainly on the right track'' \cite[pp. 103-104]{de Sitter 1932a}.

\begin{acknowledgements}
We express our gratitude to Jan Guichelaar, Roberto Lalli, Dan Lewis, Liliane Moens, Laura Rigoli, Sofia Talas, Frans van Lunteren, Luciano Vanzo. We are very grateful to Tilman Sauer for his invaluable comments and suggestions. This work has been supported in part by the Ateneo Research Projects, University of Padova.
\end{acknowledgements}

\end{document}